\documentclass[a4paper,twocolumn,showpacs,amsmath, amssymb,nofootinbib,aps,floatfix]{revtex4}
\usepackage{epsfig}
\input psfig.sty 

\usepackage{bm}									
\usepackage{dcolumn}								
\usepackage{graphicx}								
\usepackage{hyperref}

\newcommand{\no}{\mbox{$n_{e_o}$}}

\begin{document}

\title{Forecasting constraints on the cosmic duality relation with galaxy clusters}

\author{R. S. Gon\c{c}alves$^1$\footnote{E-mail: rsousa@on.br}}

\author{J. S. Alcaniz$^1$\footnote{E-mail: alcaniz@on.br}}

\author{J. C. Carvalho$^{1,2}$\footnote{E-mail: carvalho@dfte.ufrn.br}}

\author{R. F. L. Holanda$^{3,4}$\footnote{E-mail: holanda@uepb.edu.br}}

\address{$^1$Departamento de Astronomia, Observat\'orio Nacional, 20921-400, Rio de Janeiro - RJ, Brasil}

\address{$^2$Departamento de F\'{\i}sica, Universidade Federal do Rio Grande do Norte, 59072-970, Natal - RN, Brasil}

\address{$^3$Departamento de F\'{\i}sica, Universidade Estadual da Para\'{\i}ba, 58429-500, Campina Grande - PB, Brasil}

\address{$^4$Departamento de F\'{\i}sica, Universidade Federal de Campina Grande, 58429-900, Campina Grande - PB, Brasil}

\date{\today}

\begin{abstract}
One of the fundamental hypotheses in observational cosmology is the validity of the so-called cosmic distance-duality relation (CDDR). In this paper, we perform Monte Carlo simulations based on the method developed in Holanda, Gon\c{c}alves \& Alcaniz (2012) [JCAP 1206 (2012) 022] to answer the following question: what is the number of galaxy clusters observations $N_{crit}$ needed to check the validity of this relation at  a given confidence level? At $2\sigma$, we find that $N_{crit}$ should be increased at least by a factor of 5 relative to the current sample size if we assume the current observational uncertainty $\sigma_{obs}$. Reducing this latter quantity  by a factor of 2, we show that the present number of data would be already enough to check the validity of the CDDR at $2\sigma$.

\end{abstract}


\maketitle

\section{Introduction}

The variety and robustness of current astronomical data provide not only the possibility of constraining cosmological parameters but also of testing some fundamental hypotheses in Cosmology. One of these hypotheses is the validity of the so-called cosmic distance-duality relation (CDDR) (Ellis 1971, 2007), widely assumed in observational cosmology. Concisely speaking, the CDDR is derived from Etherington reciprocity theorem (Etherington 1933), which holds if photons follow null (unique) geodesic and the geodesic deviation equation is valid, along with the assumption that the number of photon is conserved over the cosmic evolution (see Bassett {{et al.}} 2004; Uzan {{et al.}} 2004; Gon\c calves {{et al.}} 2012; Ellis et al. 2013 and references therein). 

If such assumptions hold in our Universe, the luminosity ($D_L$) and angular diameter ($D_A$)  distances of sources at a given redshift $z$ are related by 
\begin{equation}
\label{CDDR}
\frac{D_L}{D_A (1+z)^2} = \eta \quad {\rm{with}} \quad \eta = 1\;.
\end{equation}
A non-validation of CDDR ($\eta \neq 1$) would be a clear evidence of new physics which could arise from different physical mechanisms. Among others, some examples are: a possible variation of fundamental constants (Jaeckel \& Ringwald, 2010), birefringence from photons (Adler 1971), a non-metric theory of gravity, a non-conservation of the number of photons due to absorption by dust (Bassett {{et al.}} 2004),  photon-axion oscillation in an external magnetic field (Cs\'aki {{et al.}} 2002) or an interactive cosmological creation process (Lima {{et al.}} 2000), among others (see, e.g., Brax et al. 2013 and references therein). 

Recently, the empirical validity of the CDDR has been explored using different methods. In some cases, a cosmological model suggested by a set of observations is assumed to place bounds on the duality parameter $\eta$ (Uzan {{et al.}} 2004; Avgoustidis {{et al.}} 2010). Cosmological model-independent tests have also been performed using $D_A$ measurements from galaxy clusters, obtained from their X-ray and Sunyaev-Zeldovich observations, and luminosity distances $D_L$ from type Ia supernovae (SNe Ia) (Holanda {{et al.}} 2010, 2011; Li {{et al.}} 2011; Nair et al. 2011; Cardone {{et al.}} 2012). In this context, Gon\c{c}alves {{et al.}} (2012) showed that X-ray measurements of the gas mass fraction of galaxy clusters (GC's) depend explicitly on the validity of Eq. (1) and discussed how these measurements  together with SNe Ia observations can be used to test the CDDR. Other cosmological model-independent tests have been proposed by using SNe Ia observations with  opacity-free distance modulus obtained from 
Hubble parameter data (Holanda {{et al.}} 2013). It is important to note that most of the methods above mentioned are potentially contaminated by different systematics error sources present in galaxy clusters, SNe Ia observations and Hubble parameter data.  

A consistent model-independent test for CDDR  that uses only observations of the gas mass fraction of GC's was proposed in Holanda, Gon\c{c}alves \& Alcaniz (2012) (from now on HGA). The basic idea is that, while current GC's gas mass fraction obtained from X-ray observations ($f_{X-ray}$) depends explicitly on the duality relation, Sunyaev-Zeldovich observations ($f_{SZE}$) of the very same quantity do not, which allows a direct comparison of $f_{X-ray}$ and $f_{SZE}$ as a test for the CDDR. Currently only 38  galaxy clusters have their gas mass fraction obtained from X-ray and microwave bands (La Roque {{et al.}} 2006). These observations provide tight but not definitive constraints on the $\eta$ parameter.

Our goal in this paper is, therefore, to answer the following question: by assuming the current observational error distribution and the respective bounds on the $\eta$ parameter, how many data points are needed to check the validity of Eq. (1) at a given confidence level?  We provide an answer to this question performing Monte Carlo simulations with different sample sizes and characteristics. We also discuss how an improvement on the $f_{X-ray}$ and $f_{SZE}$ measurements could help answer the above question. In what follows, we outline the main assumptions of our analysis and discuss the main results.

\section{$f_{gas}$ as a test of CDDR}

In order to analyze the validity of CDDR, we follow HGA and combine measurements of $f_{X-ray}$ and $f_{SZE}$ to obtain estimates of the $\eta$ parameter.

As is well known, to estimate the X-ray gas mass fraction of GC's we define a gas density profile for the electron in the core of the cluster. In general, it is assumed the model based on a spherical $\beta$-model (Cavaliere \& Fusco-Fermiano 1978) to the electron density profile. Taking the luminosity emitted from the galaxy cluster by Bremsstrahlung emission (Sarazin 1988) and assuming a hydrostatic equilibrium of the total mass of the cluster, it is possible to obtain measurements of gas mass fraction from X-ray observations (Sasaki 1996; Allen et al. 2002, 2004; Ettori et al. 2009), where the electron density of the core is written as (La Roque {{et al.}} 2006)
\begin{equation}
\label{eq:XrayNe0}
n^{X-ray}_{e0} = \left( \frac{D_L S_{x0} 4 \pi \frac{\mu_H}{\mu_e}\Gamma(3\beta)}{ D_A^3 \: \Lambda_{eH} \: \pi^{1/2}  \Gamma(3\beta-\frac{1}{2}) \theta_c} \right)^{1/2}\; .
\end{equation}
In the above expression, $S_{x0}$, $\mu_H$, $\mu_e$, $\beta$, $\Lambda_{eH}$ and $\theta_c$ are, respectively, the flux in x-ray, the mean molecular weight of Hydrogen, the mean molecular weight of electrons, the power law index from $\beta$ model, the X-ray cooling function and the angular diameter of the core. From Eq. (1) and (2), it is clear that the density of electron in the core depends on the CDDR parameter since it depends explicitly on the ratio $D_L/D_A$. Consequently, the gas mass fraction obtained via X-ray measurements is affected by a possible violation of CDDR (Gon\c calves {{et al.}} 2012).

\begin{figure}
\psfig{figure=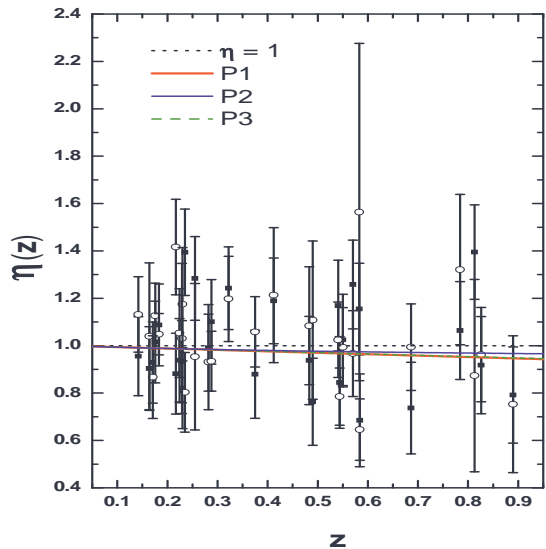,width=3.0in,height=3.0in}
\caption{$\eta(z)$ as a function of $z$ obtained from the La Roque {{et al.}} sample of $f_{X-ray}$ and $f_{SZE}$ data (open circles). The curves represent the best-fit for each parametrization adopted in the analysis (see Eq. \ref{ps}). For comparison, one Monte Carlo realization of 29 values of the CDDR parameter (filled squares) is also shown.}
\end{figure}

On the other hand, combining measurements of the decrement of the CMB temperature (Sunyaev \& Zel'dovich 1972) as it propagates through the intracluster medium with the shape of the X-ray spectra provides $f_{SZE}$ measurements. In this case (La Roque {{et al.}} 2006), 
\begin{equation}
\label{eq:SZNe0}
n_{e0}^{SZE} = \left( \frac{\Delta To \, m_e c^2 \: \Gamma(\frac{3}{2}\beta)}{D_A \: f_{(\nu, T_e)}
  T_{cmb} \sigma T \, k_{\rm B} T_e \pi^{1/2} \: \Gamma(\frac{3}{2}\beta - \frac{1}{2})\, \theta_c} \right),
\end{equation}
where the parameters are $\Delta To$ (the variation of temperature in the core), $m_e$ (electron mass), $f_{(\nu, T_e)}$ (a function that accounts for frequency shift and relativistic corrections), $T_{cmb}$ (CMB temperature), $k_{\rm B}$ (Boltzmann constant) and $T_e$ (gas temperature). Measurements of gas mass fraction from the Sunyaev-Zel'dovich effect do not have a clear dependence on the CDDR\footnote{As mentioned in HGA, $f_{SZE}$ measurements are redshift-independent only if there is no process of energy injection into the CMB. In cosmologies with photon creation or destruction, the standard linear relation $T(z) = T_0(1+z)$ is not valid and changes are needed in Eq. (\ref{eq:SZNe0}).}. Since current $f_{X-ray}$ measurements have been obtained by assuming the validity of the CDDR,  a direct combinations of the above equations with other observational quantities provides a general expression relating current X-ray and SZE observations, namely,
\begin{equation}
\label{method}
f_{SZE} = \eta f_{X-ray} \; \; .
\end{equation}
We refer the reader to HGA for a complete derivation of the above expression and a detailed discussion.

\begin{figure*}
\psfig{figure=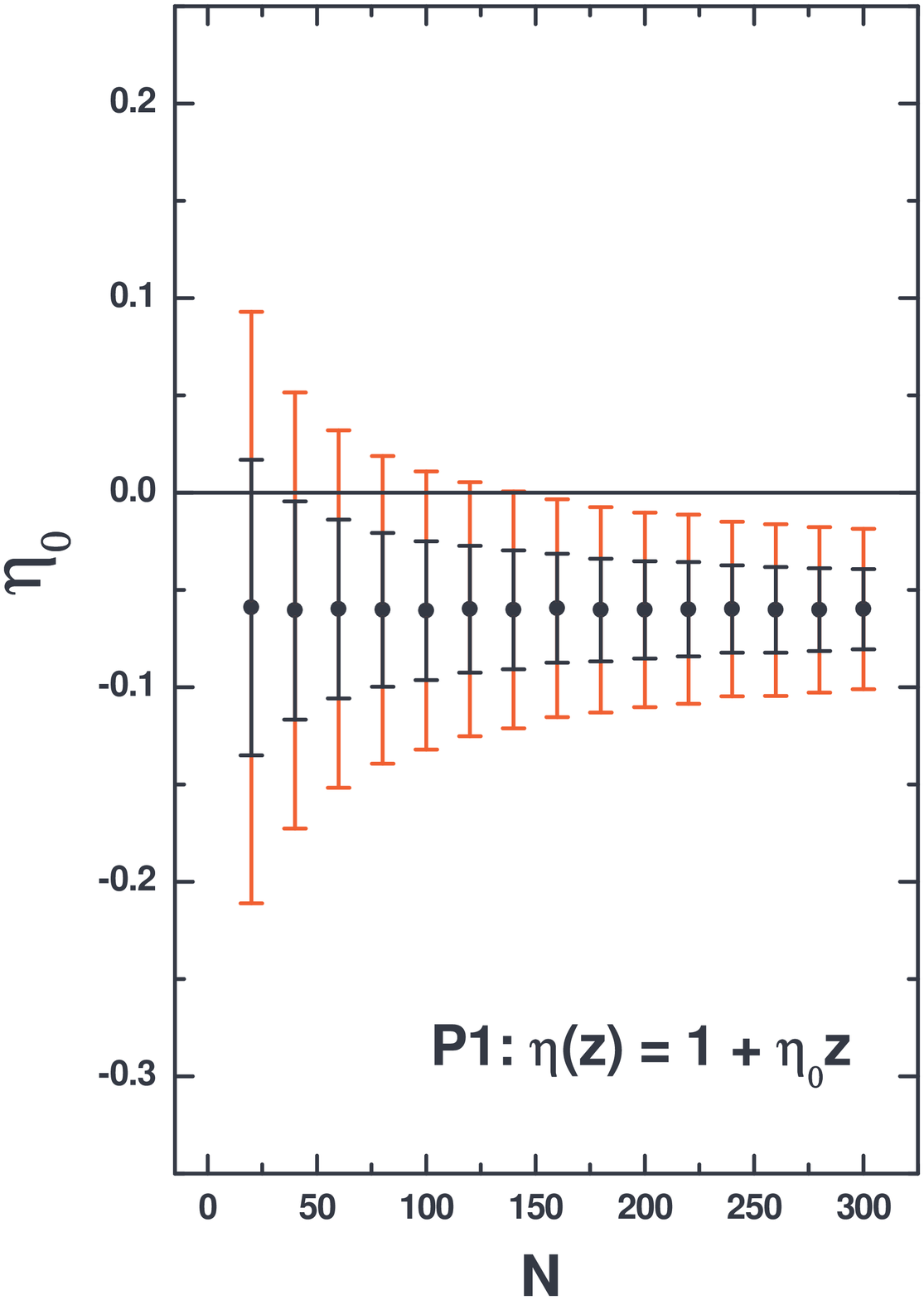,width=2.3in,height=2.3in}
\psfig{figure=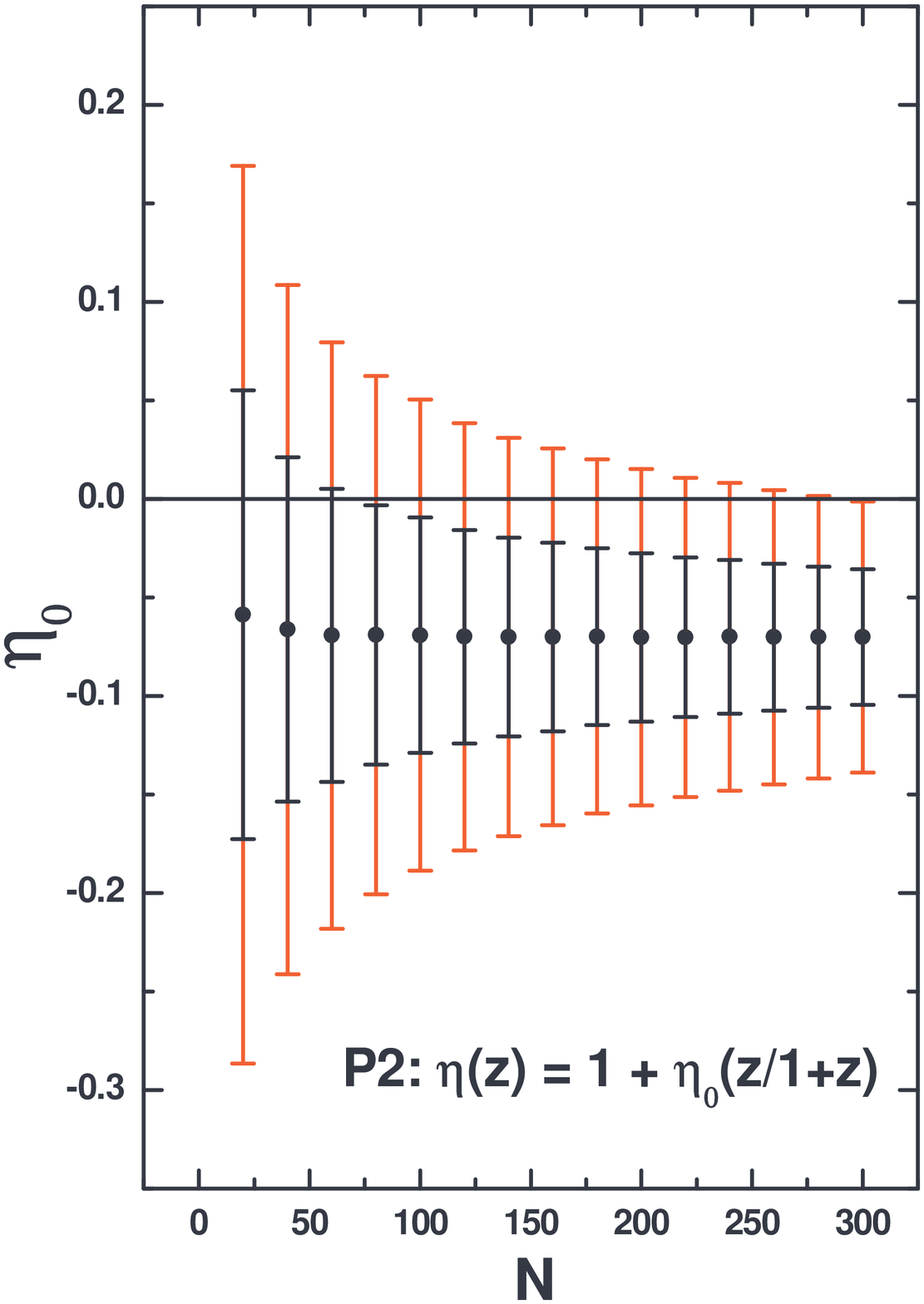,width=2.3in,height=2.3in}
\psfig{figure=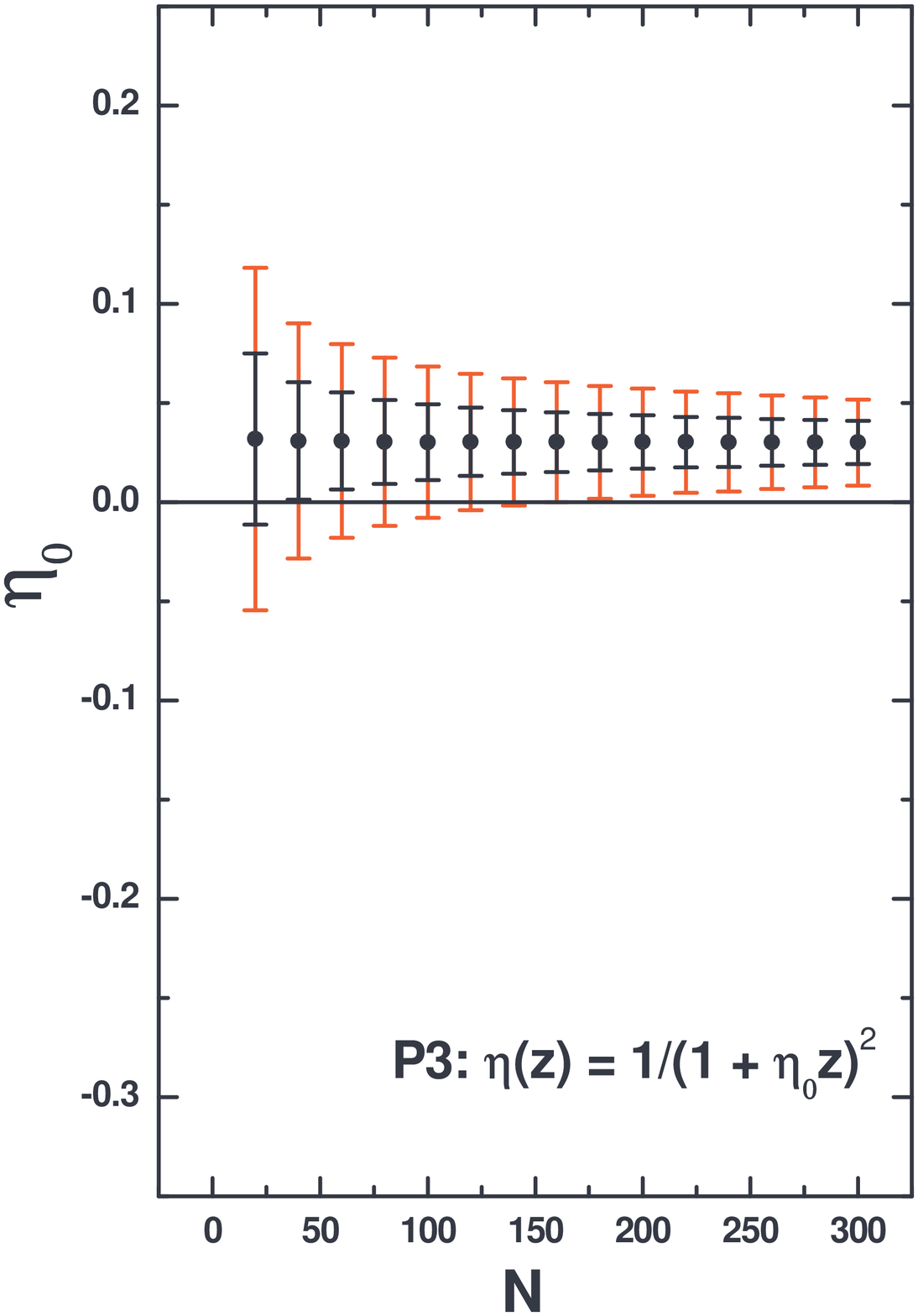,width=2.3in,height=2.3in}
\caption{The parameter $\eta_0$ as a function of the number of data points $N$ for the three parameterizations discussed in the text. The error bars correspond to $1\sigma$ (black) and $2\sigma$ (red). The value of $N_{crit}$, obtained from Eq. (\ref{criterion}), depends strongly on the $\eta(z)$ parameterization adopted.}
\end{figure*} 

\section{Observational Data}

In our analysis, we use a subset of $f_{gas}$ data provided by La Roque { et al.} (2006). Our sample is composed of $29$ measurements  obtained via X-ray and SZE observations.  The X-ray data were obtained from the Chandra X-ray Observatory and SZE data from the BIMA/OVRO SZE imaging project, which uses the Berkeley-Illinois-Maryland Association (BIMA) and Owens Valley Radio Observatory (OVRO) interferometers to image the SZE. The original data set contains 38 data points spanning redshifts from $0.14$ up to $0.89$, but we excluded 9 points that do not have a good agreement with a smooth description of their density profile and with the hydrostatic equilibrium model. The galaxy clusters excluded from our sample are {\it{Abell 665}}, {\it{ZW 3146}}, {\it{RX J1347.5-1145}}, {\it{MS 1358.4 + 6245}}, {\it{Abell 1835}}, {\it{MACS J1423+2404}}, {\it{Abell 1914}}, {\it{Abell 2163}}, {\it{Abell 2204}} whose observations present a reduced $\chi^2$ ranging between $2.43 \leq  \chi^2 \leq  41.62$. A detailed 
discussion about the statistical and systematic errors of these X-ray and SZE observations can be found in Bonamente {{et al.}} (2006).

As commented by HGA, $f_{gas}$ obtained via SZE is not completely independent of X-ray observations, since the shape parameters of the gas density model (core radius and $\beta$ parameter) are obtained from a joint analysis of the X-ray and SZE data (Bonamente {{et al.}} 2004). However, current simulations (Hallman {{et al.}} 2007) have  shown that such parameters, when obtained separately by SZE and X-ray observations, agrees at 1 $\sigma$ level within a radius $r_{2500}$ (at which the mean enclosed mass density is equal to 2500 cosmological critical density), the same used in the La Roque {{et al.}} (2006) observations. Therefore, one can use the La Roque {{et al.}} (2006) sample to perform analysis with the method discussed earlier. 

In order to run our simulations and perform the statistical analyses in the next section, we need to define a proper parametrization for $\eta$. It is worth noting that any parametrization adopted must satisfy the condition $\eta(z \sim 0) \simeq 1$ since at very low-$z$ $D_L = D_A = H_0^{-1}z + O(z^2)$. In our analysis, we consider the following three parameterizations:
\begin{eqnarray}
\label{ps}
\eta = \eta(z) = \; \left\{
\begin{tabular}{l}
$1 + \eta_0 z$ \quad  \quad  \quad  \quad   \hspace{0.45cm}(P1)  \\
\\
$1 + \eta_0\frac{z}{1+z}$ \quad \quad \quad \hspace{0.3cm} (P2)\\
\\
$1/(1 + \eta_0 z)^2$ \quad  \quad  \hspace{0.38cm}(P3) \quad 

\end{tabular}
\right.
\end{eqnarray}
Figure 1 shows the CDDR parameter $\eta$ as a function of $z$ obtained from current $f_{X-ray}$ and $f_{SZE}$ observations (open circles). The best-fit curves for parameterizations P1, P2 and P3 are also shown.

\begin{figure*}
\psfig{figure=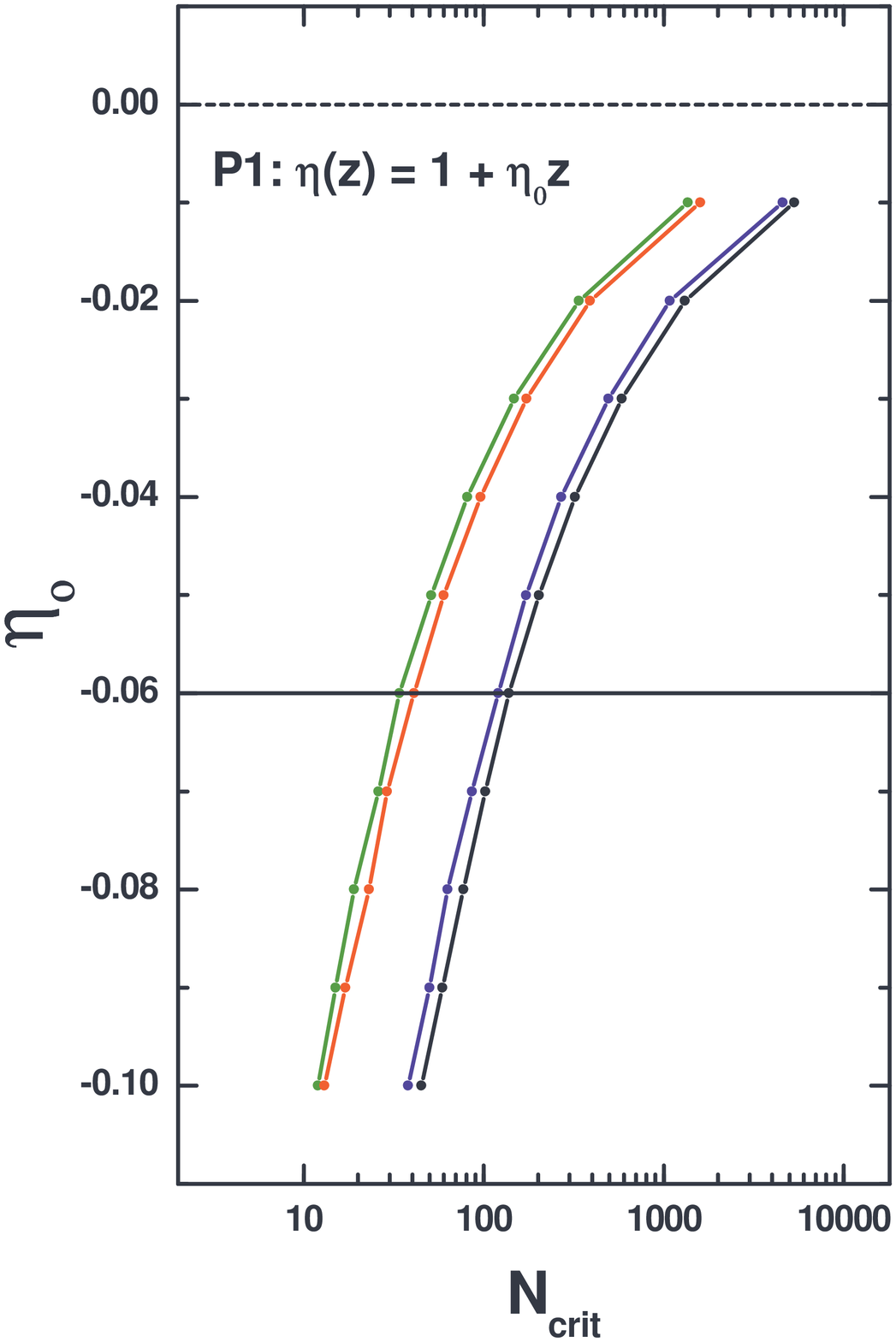,width=2.3in,height=2.5in}
\psfig{figure=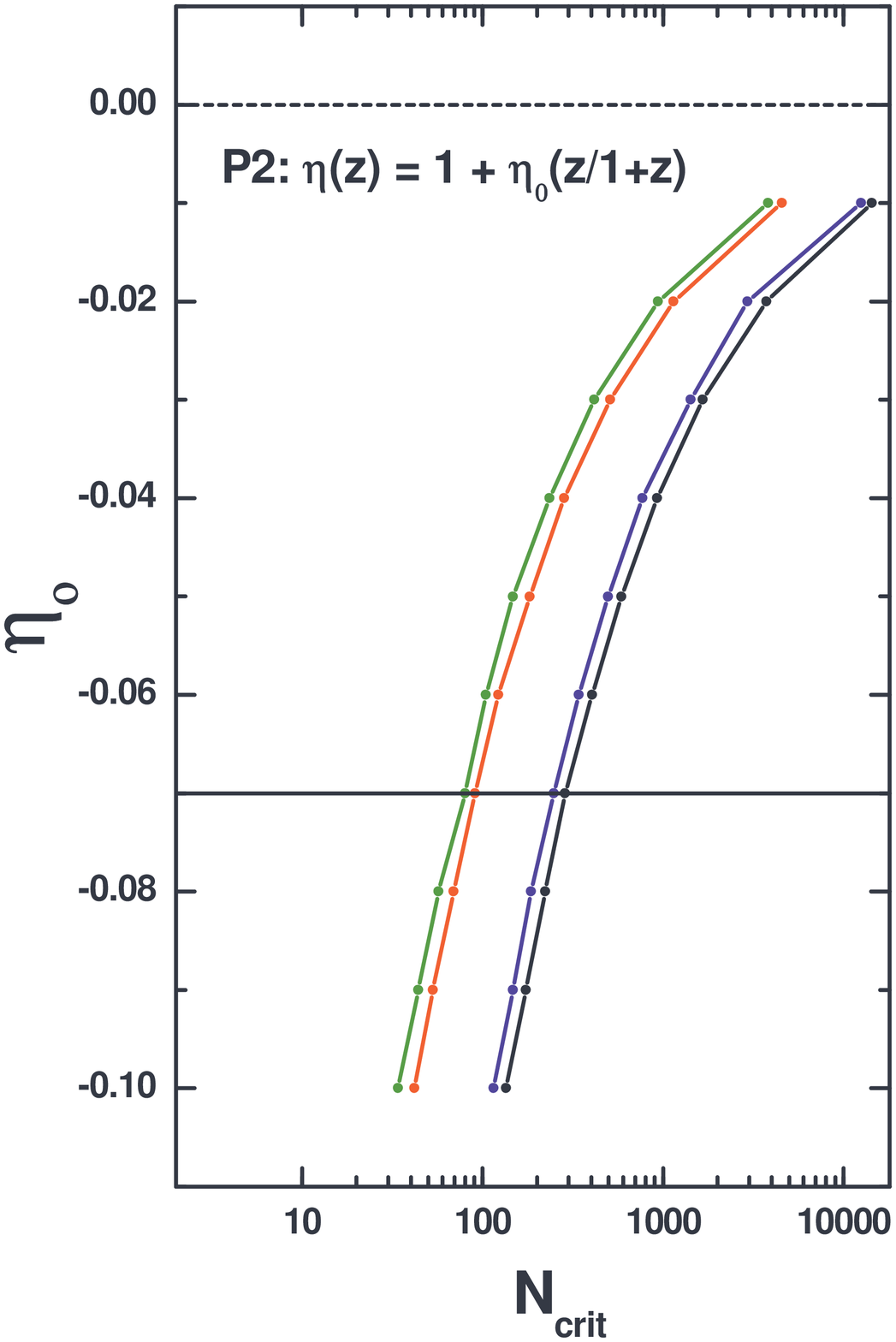,width=2.3in,height=2.5in}
\psfig{figure=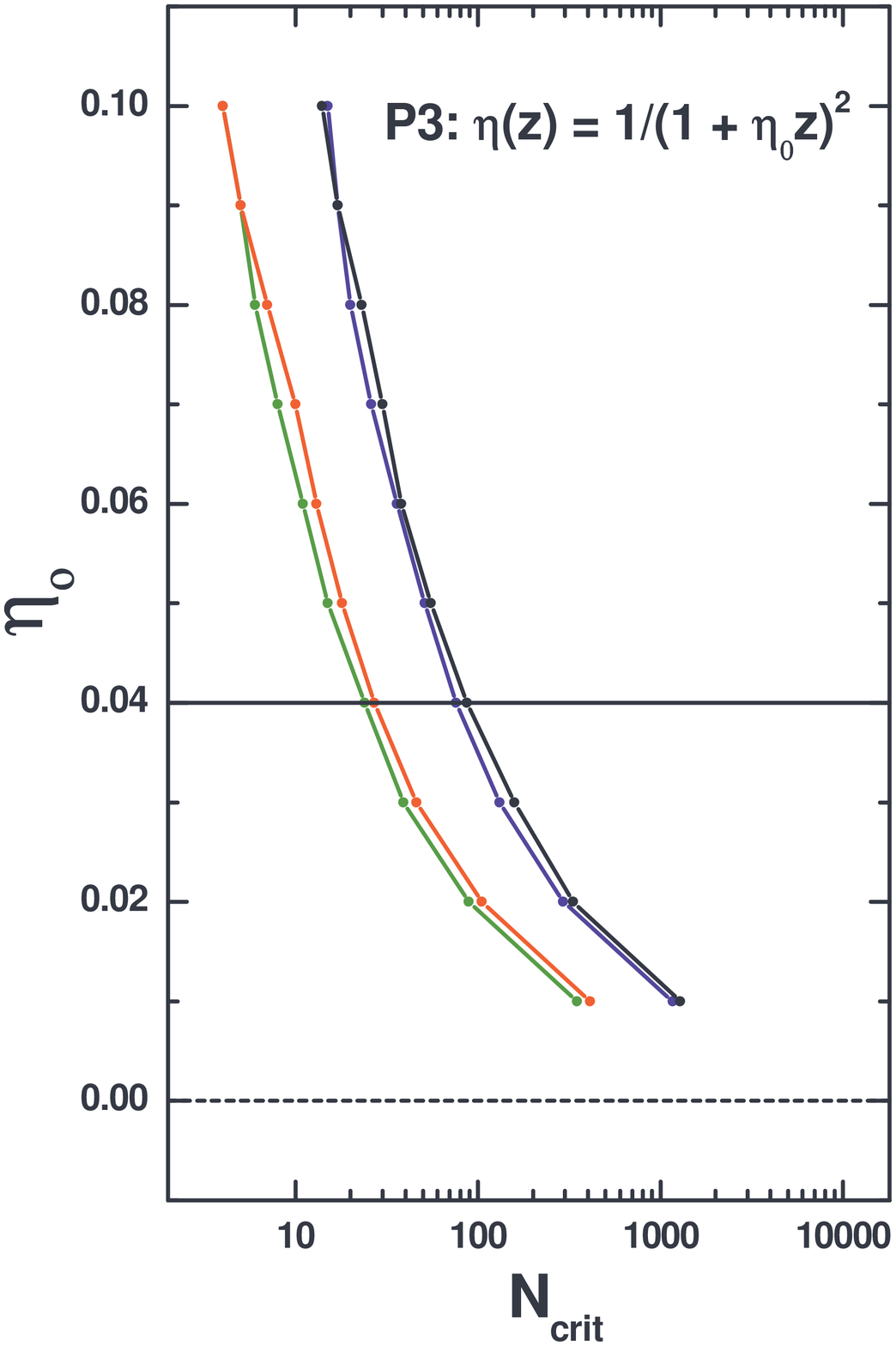,width=2.3in,height=2.5in}
\caption{The parameter $\eta_0$ as a function of $N_{crit}$ for parameterizations P1 (left), P2 (middle) and P3 (right). As discussed in the text, the black curves are obtained assuming the current observational uncertainty on $\sigma_{X-ray}$ and $\sigma_{SZE}$ whereas the dark blue ones are obtained fixing $\sigma_{SZE}$ and assuming $\sigma_{X-ray}/2$. The case $\sigma_{SZE}/2$ and $\sigma_{X-ray}$ is represented by the red lines. Green lines represent a reduction of $50\%$ on both $\sigma_{X-ray}$ and $\sigma_{SZE}$ relative to the current data.}
\end{figure*}

\section{Simulation}

The current $f_{gas}$ observations do not provide a definite answer about the validity of the CDDR (Gon\c calves, Holanda \& Alcaniz 2012). However, using the method described above [see Eq. (\ref{method})], it is expected that upcoming observations with larger number and more precise data will be able to test the CDDR more accurately. In what follows, by assuming the observational uncertainty of the current $f_{gas}$ observations, we discuss how large should the upcoming $f_{gas}$ samples be in order to check the validity of the CDDR at a given confidence level.

To this end, we run Monte Carlo simulations and generate synthetic samples of $\eta(z)$ measurements in the redshift interval $[0.1, 1.0]$. The process of simulation can be described as follows. To begin with, we assume a parametrization for $\eta(z)$ and make a $\chi^2$ analysis to obtain the best fit value of $\eta_0$ from the observational data set. We also calculate the relative error ($\sigma_{\eta}/\eta$) and fit a straight line in order to find its dependence on redshift. 

The simulated sample of $\eta(z)$ is obtained by drawing $N$ random values of $\eta_0$ from a normal distribution centered at the fitted value of $\eta_0$ and with standard deviation equal to the data dispersion. The points are equally spaced in the redshift interval [0.1, 1]. Similarly, the error bars for each simulated point are drawn from a normal distribution with a mean value equal to the dispersion of the real data and taking into account the previously found redshift dependence. The next step is to make a $\chi^2$ statistical analysis of the synthetic data (with respective errors)  and obtain the best fit value of $\eta_0$ for each sample. We repeat this process $10^4$ times and calculate the mean and standard deviation for this group of best fit. The process is repeated for different values of $N$ in order to study the effect of sample size on the determination of $\eta_0$. For the sake of illustration, one Monte Carlo realization of 29 values of the CDDR parameter (filled squares) is shown in Fig. 
1.

\section{Results}

The simulation process above provides the number of data points $N_{crit}$ necessary to check the validity of the CDDR for a specific parametrization, with a fiducial value of $\eta_0$. In order to find $N_{crit}$, we adopt the criterion 
\begin{equation}
\label{criterion}
n\sigma_{\eta_0}/\eta_0 < 1\; ,
\end{equation}
where $n$ corresponds to the confidence level required. For the $\eta(z)$ parameterizations discussed earlier [Eqs. (\ref{ps})], we find the following results for $n = 2$ ($2 \sigma$) (see Figs. 2a-2c):

$$
{\rm{P1:}}  \hspace{0.45cm} \eta_0  = -0.06 \hspace{0.45cm}  N_{crit} = 138;\\
$$
$$
{\rm{P2:}}  \hspace{0.45cm} \eta_0  = -0.07 \hspace{0.45cm}  N_{crit} = 287;
$$
$$
{\rm{P3:}}  \hspace{0.46cm} \eta_0 = +0.04 \hspace{0.46cm}  N_{crit} = 158.
$$
which shows a clear dependence of $N_{crit}$ with the parameterization adopted. For completeness, we also perform simulations for different fiducial values of $\eta_0$, i.e., changing this value in our simulation process. As expected, the closer to zero this latter value the larger $N_{crit}$. These results are shown in Figures 3a-3c (black lines) for parameterizations P1, P2 and P3, respectively.

It is worth noticing that the above results are somehow conservative since no improvement on the upcoming $f_{gas}$ data was considered to derive the above values of $N_{crit}$. Therefore, an interesting question worth asking is: what is the influence of the uncertainties $\sigma_{X-ray}$ and $\sigma_{SZE}$ on the estimates of $N_{crit}$? The answer to this question is also shown in Figures 3a-3c for parameterization P1, P2 and P3, respectively. Blue lines represent a reduction of $50\%$ on $\sigma_{X-ray}$ relative to the La Roque {{et al.}} sample whereas red lines represent a reduction of $50\%$ on $\sigma_{SZE}$.  A reduction of $50\%$ on both $\sigma_{X-ray}$ and $\sigma_{SZE}$ from the current data is also shown (green lines). Clearly, $f_{SZE}$ observations are the major source of uncertainty to test the CDDR from the method described in Sec. 2. According to (\ref{criterion}), we also note that reducing the current uncertainty on $\sigma_{X-ray}$ and $\sigma_{SZE}$ by $50\%$ the current number of data 
points 
would be already enough to check the validity of the CDDR at 2$\sigma$ level.

\section{Conclusion}

Testing the CDDR constitutes an important task for cosmology and fundamental physics. In this paper, we have adopted the method proposed in HGA (Sec. II) and forecasted constraints on the CDDR parameter from Monte Carlo simulations.  

Initially, we have assumed the present observational error distribution, generated samples of gas mass fraction in the redsfhit interval $[0.1, 1]$ and derive $\eta(z)$ values using a direct relation between $f_{X-ray}$ and $f_{SZE}$ [Eq. (\ref{method})]. For the three parameterizations adopted in our analysis, we have found that the minimum number of data points necessary to validate the CDDR at 2$\sigma$ level should be from 5 to 10 times larger than the current observational sample. 

We have also analyzed the influence of the observational errors in the $f_{X-ray}$ and $f_{SZE}$ observations on our results. Regardless of the parameterization adopted, we have found that the major source of uncertainty comes from $f_{SZE}$ measurements, with $N_{crit}$ being reduced by a factor of 4 when $\sigma_{SZE}$ decreases by half. These results clearly show that a combination between quantity and precision of future $f_{gas}$ measurements can become a fundamental tool to check the validity of the CDDR and explore its consequences.

\acknowledgements

The authors thank CNPq, INCT-A, INEspa\c{c}o and FAPERJ for the grants under which this work was carried out.

\end{document}